\documentclass[prl,twocolumn,groupedaddress,showpacs,showkeys,amsfonts]{revtex4}

\newcommand{\tr}{{\textrm {tr}}}
\newcommand{\Tr}{{\textrm {Tr}}}

\newcommand{\D}{{\mathcal D}}
\newcommand{\ha}{a^{T}}
\newcommand{\hb}{b^{T}}

\newcommand{\bfx}{\mbox{\boldmath $x$}}

\begin{document}

\title{The Polyakov loop and the heat kernel expansion at finite temperature}

\author{E. Meg\'{\i}as}

\author{E. \surname{Ruiz Arriola}}

\author{L.L. Salcedo}

\affiliation{
Departamento de F\'{\i}sica Moderna,
Universidad de Granada,
E-18071 Granada, Spain
}

\date{\today} 

\begin{abstract}
The lower order terms of the heat kernel expansion at coincident
points are computed in the context of finite temperature quantum field
theory for flat space-time and in the presence of general gauge and
scalar fields which may be non Abelian and non stationary. The
computation is carried out in the imaginary time formalism and the
result is fully consistent with invariance under topologically large
and small gauge transformations. The Polyakov loop is shown to play a
fundamental role.
\end{abstract}



\maketitle

The heat kernel $e^{-\tau K}$, where $K$ is an Euclidean Klein-Gordon
operator, was introduced by Schwinger \cite{Schwinger:1951nm} in
quantum field theory as a tool to regularize ultraviolet divergences
in a gauge invariant way. The procedure of systematically isolating
these divergences was developed by DeWitt through an asymptotic
expansion in powers of the proper time $\tau$
\cite{Dewitt:1975ys,Seeley:1967ea}. The heat kernel as well as its
expansion has also been applied to study spectral densities and
indices of Dirac operators ($\hat D$)
\cite{Gilkey:1975iq,Atiyah:1973ad} in terms of effective Klein-Gordon
operators ($\hat D^\dagger\hat D$), to compute the $\zeta$-function
\cite{Hawking:1977ja} and anomalies of such operators
\cite{Fujikawa:1980eg}, to properly define the effective action of
chiral gauge theories \cite{Ball:1989xg}, to the Casimir effect
\cite{Bordag:2001qi}, etc. Exact calculations of the heat kernel at
coincident points are available in particular manifolds
\cite{Bytsenko:1996bc,Camporesi:1990wm}, and the coefficients of the
expansion have been computed to rather high orders in several setups,
including curved spaces with and without boundary, and in presence of
non Abelian gauge fields and non Abelian scalar fields, using
different methods
\cite{Ball:1989xg,Bel'kov:1996tn,vandeVen:1998pf,Moss:1999wq,%
Fliegner:1998rk,Avramidi:1991je,Gusynin:1989ky}. In this work we deal
with the extension of these results to finite temperature in flat
space-time.

As is well known, in the Matsubara formalism for quantum field theory
at finite temperature, wave functions are required to be periodic for
bosons and antiperiodic for fermions, with period $\beta$ (the inverse
temperature) and the external fields are also periodic. This
introduces a compactified Euclidean time \cite{Landsman:1987uw}. The
heat kernel can be regarded as the evolution operator (in imaginary
proper time $\tau$) of a first quantized particle
\cite{Schwinger:1951nm}, thus it admits a path integral
representation, which at finite temperature must include also multiply
connected paths in the compactified space-time.  The standard results
quoted e.g. in \cite{Ball:1989xg} for the heat kernel expansion refer
to general smooth manifolds without boundary and thus, in principle,
they would directly apply to finite temperature as well. However, the
standard formulas are not sensitive to the compactification of time;
they only sum paths which are topologically trivial (contractile to a
point). Actually this correctly reproduces the analytical part in a
strict expansion in powers of the proper time, but neglects the
exponentially suppressed finite temperature contributions
\cite{Actor:1994xq}. To see any temperature effect in the heat kernel
expansion it is necessary to include also the paths that wind once or
more around the compactified time. This is automatically achieved in
the canonical formalism by replacing the continuous energy variable by
the discrete Matsubara frequencies.  More generally, as already argued
in \cite{Garcia-Recio:2000gt} (see also \cite{Deser:2002vx}), in the
presence of gauge fields, finite temperature effects are tied to the
presence of the (untraced) Polyakov loop
\begin{equation}
\Omega(x)= T\exp\left(-\int_{x_0}^{x_0+\beta}
A_0(x_0^\prime,\bfx)dx_0^\prime\right)
\end{equation}
($T$ refers to temporal ordering, this is required to allow for the
non Abelian case). The presence of $\Omega(x)$ is to be expected on
general grounds. Let us regard the heat kernel at finite temperature
as the evolution operator of a first quantized particle. The
space-time manifold defines a $d$-dimensional cylinder which plays of
role of space for the particle and $-i\tau$ is the time. If the
circulation of the ``vector potential'' $A_\mu$ around the cylinder is
non vanishing this signals the presence of an Aharanov-Bohm magnetic
flux through its interior, which will affect the propagation. The
particle will pick up a phase $\Omega(x)$ after looping once around
the cylinder (in the general case a non Abelian phase
\cite{Wu:1975vq}).

A computation of the thermal heat kernel coefficients was presented in
\cite{Boschi-Filho:1992ah} where a common renormalization for all the
coefficients, $\ha_n=f(\tau/\beta^2) a_n$, was found.  However,
because the temperature effects account for the fact that Euclidean
time is compact, an explicit breaking of Lorentz invariance is
expected and the previous factorization cannot hold in general (see
e.g. \cite{Actor:2000hc}).  The former computation refers to flat
space-time (without boundaries) and non Abelian scalar and gauge
fields. In \cite{Xu:1993qi} a similar study is presented where
curvature is allowed but fields are assumed to be static. These
authors find that $f(\tau/\beta^2) a_n$ is augmented with new terms
that break Lorentz symmetry, so-called anomalous terms also found in
\cite{Nakazawa:1985zq}. However both calculations are incomplete since
they are in conflict with results for Abelian and stationary
configurations (in flat space) presented in \cite{Actor:2000hc}. There
it is shown that the Euclidean scalar potential $A_0(\bfx)$ can appear
in the heat kernel (outside covariant derivatives) provided that it
does so as a periodic variable with period $2\pi/\beta$, to be
consistent with (large) gauge invariance. This is a particular
manifestation of the relevance of the Polyakov loop.

We present here the first systematic calculation of the heat kernel
expansion at finite temperature at coincident points (in flat space
without boundaries) for general backgrounds which may be non Abelian
and non stationary. The result is gauge covariant and depends locally
on the covariant derivatives and the Polyakov loop. The Klein-Gordon
operator to be considered is of the type $K=M-D_\mu^2$, where
$D_\mu=\partial_\mu+A_\mu(x)$ is the gauge covariant derivative and
$M(x)$ is a Lorentz scalar. Both $A_\mu(x)$ and $M(x)$ are allowed to
be matrices in some internal space (the treatment applies to Abelian
and non Abelian theories). One can form local and gauge covariant
operators by combining the basic operators $M$,
$F_{\mu\nu}=[D_\mu,D_\nu]$ and their covariant derivatives, e.g.
${\cal O} = F_{\mu\nu}\D_\mu \D_\nu M $, where $\D_\mu X$ stands for
$[D_\mu,X]$. The local operators can be classified according to their
mass dimension. For each dimension there is a finite number of
them. At zero temperature, under suitable regularity conditions, one
can asymptotically expand an expression of the form $\langle x|f(K)|
x\rangle$ as a sum of local operators, i.e.
\begin{equation}
\langle x|f(K)| x\rangle= \sum_r f_r {\cal O}_r(x) \,,
\end{equation}
${\cal O}_r$ denotes a linear basis including all possible local
operators and $f_r$ are some numerical coefficients which depend on
the function $f$. In particular, for the matrix elements of the heat
kernel at coincident points one obtains the familiar Seeley-DeWitt
asymptotic expansion
\begin{equation}
\langle x|e^{-\tau K}| x\rangle= (4\pi\tau)^{-d/2}\sum_{n=0}^\infty
a_n(x)\tau^n \,,
\end{equation}
$d$ being the space-time dimension. The local operators $a_n(x)$
themselves do not depend on $d$ or $\tau$ and have mass dimension
$2n$. Of course, once one has the expansion for the heat kernel, the
use of integral representations \cite{Schwinger:1951nm} allows to
obtain the expansion for other functions $f(K)$, such as the
$\zeta$-function \cite{Hawking:1977ja,Seeley:1967ea}, the effective
action, the resolvent or the effective current, without a new
computation.

At finite temperature, because the Euclidean time coordinate is
compactified, one can expect to expand $\langle x|f(K)| x\rangle$ in
terms of operators which are local regarding the spatial coordinates
but not in the time coordinate \cite{Dunne:1997yb}. Nevertheless, the
following asymptotic expansion still holds \cite{Garcia-Recio:2000gt}
\begin{equation}
\langle x|f(K)| x\rangle= \sum_r f_r(\Omega(x)) {\cal O}_r(x) \,.
\label{eq:4}
\end{equation}
That is, at finite temperature the numerical coefficients $f_r$ become
functions of the Polyakov loop $\Omega(x)$. This result is remarkable
on several accounts. First, although full space-time locality does not
hold at finite temperature, still the result looks space-time local if
written in terms of the field $\Omega(x)$ and the other local
operators. (Note that $\Omega(x)$ is local regarding gauge
transformations.) Second, full gauge covariance (the covariance under
continuous local similarity transformations in internal space) is
manifest, since this property is shared by the Polyakov loop and the
local operators. This is not always automatically true for other
expansions; because the space-time is not simply connected in the
imaginary time formalism, in addition to the usual topologically small
gauge transformations there are also large transformations (i.e., not
reachable by composition of infinitesimal ones). Covariance under such
large transformations fails to hold, for instance, in simple-minded
perturbation theory \cite{Dunne:1997yb,Salcedo:2002pr}.

In the non Abelian case $\Omega(x)$ needs not commute with the other
quantities. The fact that it appears to the left in (\ref{eq:4}) is a
matter of choice (putting it to the right would yield a different set
of functions $f_r$). We remark that the operators ${\cal O}_r$ are
local and gauge covariant but not necessarily Lorentz invariant. That
is, operators with different Lorentz indices may have different
weights at finite temperature. Rotational invariance is of course
preserved.

As we have argued above, for many purposes it is sufficient to
consider the expansion for the heat kernel. At finite temperature the
result can still be written as
\begin{equation}
\langle x|e^{-\tau K}| x\rangle= (4\pi\tau)^{-d/2}\sum_n
\ha_n(x)\tau^n \,.
\end{equation}
The thermal heat kernel coefficients $\ha_n$ contain local operators,
and depend on $\Omega$, $\tau$ and $\beta$, but not on $d$. 

At finite temperature the heat kernel expansion consists actually of
three expansions. One in powers of $M$, another in powers of $\D_i$,
and finally another expansion in powers of $\D_0$. The first two
expansions are standard and no different to those done at zero
temperature, but the expansion in $\D_0$ is tricky and has to be
carried out with care. This is because a naive expansion in temporal
derivatives is in conflict with gauge invariance. This is easy to
understand by noting that the derivative expansion refers to external
fields with very wide profiles. Whereas this is unproblematic in the
spatial direction, the situation in the temporal direction is
different since the Euclidean time is periodic and dilatations in the
temporal direction break the periodicity condition.

For the calculation we use the method of symbols which is
based on the identity \cite{Salcedo:1996qy,Salcedo:1998sv}
\begin{equation}
\langle x| f(D_\mu,M)|x\rangle =
\frac{1}{\beta}\sum_{p_0}\int \frac{d^{d-1} p}{(2\pi)^{d-1}} 
\langle x|f(D_\mu+ip_\mu,M)| 0\rangle \,.
\end{equation}
$f(D_\mu,M)$ is any (ultraviolet finite) operator constructed out of
$D_\mu$ and $M$, and $|0\rangle$ is the zero momentum state, $\langle
x|0\rangle=1$. In practice this means that $\partial_\mu$ in $D_\mu$
vanishes when it reaches the right end, and the brackets $\langle
x|\cdot|0\rangle$ are often not explicit. An application of the
symbols method formula in the temporal direction only, gives
\begin{eqnarray}
\langle x|e^{-\tau K}| x\rangle &=& 
\frac{1}{\beta}\sum_{p_0}
\langle \bfx|e^{-\tau(M-Q^2-D_i^2)}|\bfx\rangle \,,
\nonumber \\
Q &=& ip_0 + D_0 \,.
\label{eq:11}
\end{eqnarray}
For the spatial part we can directly use the known heat kernel
expansion at zero temperature \cite{Ball:1989xg,Bel'kov:1996tn} on the
effective $(d-1)$-dimensional Klein-Gordon operator
$(M-Q^2)-D_i^2$. The operators $Q$ are then moved to the left
generating commutators of the form $[Q,\ ]$. These commutators are
just $\D_0=[D_0,\ ]$ since $p_0$ is a c-number. The $Q$'s not in
commutators, placed at the left of the expressions so obtained, come
in the form $Q^n e^{\tau Q^2}$ (cf. (\ref{eq:7})). The final step is
to use the operator identity \cite{Garcia-Recio:2000gt}
\begin{equation}
e^{-\beta D_0} = e^{-\beta \partial_0}\Omega(x) \,.
\end{equation}
Because all external fields are periodic, $e^{-\beta \partial_0}=1$
and $e^{-\beta D_0}$ is equivalent to $\Omega(x)$. In this way the
operator $Q$ defined in (\ref{eq:11}) reduces to a function of $x$
\begin{equation}
Q= ip_0-\frac{1}{\beta}\log(\Omega) \,.
\label{eq:11a}
\end{equation}

An explicit computation gives, up to operators of mass dimension four
\footnote{Further details, as well as the rather long coefficients
$\ha_{5/2}$ and $\ha_{3}$, will be presented elsewhere
\cite{Megias:2003prep}.},
\begin{eqnarray}
\ha_0 &=& \varphi_0 \nonumber \\
\ha_1 &=& -\varphi_0 M
\\
\ha_{3/2} &=& \varphi_1 \left(M_0-\frac{1}{3}E_{ii}\right)
\nonumber  \\
\ha_2 &=& 
-\left(\frac{1}{2}\varphi_0+\frac{2}{3}\varphi_2\right) M_{00}
\nonumber \\ && 
+\left(\frac{1}{6}\varphi_0+\frac{1}{3}\varphi_2\right) E_{0ii}
+\left(\frac{1}{3}\varphi_0+\frac{1}{3}\varphi_2\right) E_i^2 
\nonumber \\ && 
+\varphi_0\left(\frac{1}{2}M^2-\frac{1}{6}M_{ii}+\frac{1}{12}F_{ij}^2\right) \,.
\nonumber
\end{eqnarray}
In these formulas $E_i=F_{0i}$ is the electric field and we use the
convenient notation $M_{\mu\nu}=\D_\mu\D_\nu M$, $E_{\mu i}= \D_\mu
E_i$, etc. The quantities $\varphi_n$ are functions of $\Omega(x)$ and
the ratio $\tau/\beta^2$, namely 
\begin{eqnarray}
\varphi_n(\Omega) &=&
\frac{(4\pi\tau)^{1/2}}{\beta}\sum_{p_0} \tau^{n/2} Q^n
e^{\tau Q^2} \,.
\label{eq:7}
\end{eqnarray}
Here, $Q$ given in (\ref{eq:11a}) and $p_0$ runs over the Matsubara
frequencies, $2\pi k/\beta$ ($k$ integer) for bosons and $2\pi
(k+\frac{1}{2})/\beta$ for fermions.  Consequently, there are bosonic
and fermionic versions for each function $\varphi_n$, related through
the replacement $\Omega \to -\Omega$. The $\varphi_n$'s are
related to the special functions $\theta_3$ and $\theta_4$ and their
derivatives \cite{Abramowitz:1970bk}.

The first thing to note is that, due to the summation over $p_0$, the
$\varphi_n(\Omega)$'s are one-valued functions of $\Omega$ and so
gauge invariance is preserved. (The quantity $\log\Omega$ is covariant
under small gauge transformations but many-valued under large ones.)
Actually, if we use Poisson's summation formula for, say, the function
$\varphi_0$ we obtain
\begin{equation}
\varphi_0(\Omega)= \sum_{k\in \mathbb{Z}} \Omega^k e^{-k^2\beta^2/4\tau}
\quad \text{(bosonic)} \,.
\label{eq:8}
\end{equation}
Second, we note that the expansion contains coefficients of
half-integer order with local operators of odd dimension. Due to the
property $\varphi_n(\Omega)= (-1)^n\varphi_n(\Omega^{-1})$ the
symmetry under $D_0\to -D_0$, which is manifest in the Klein-Gordon
operator, is preserved. Odd dimensional terms vanish at zero
temperature except in presence of boundaries \cite{McAvity:1991we}.
One of the techniques to deal with boundaries is the method of
images. The path integral method suggests that the presence of these
terms at finite temperature can also be understood from the fact that
the particle sees its own image from its periodic replicas.

In the zero temperature limit the functions $\varphi_n$ become
constant. For even $n$, they take the values
$(-\frac{1}{2})^{n/2}(n-1)!!$ and vanish for odd $n$. It is easy to
verify that the thermal coefficients reproduce the zero temperature
ones in this limit and Lorentz invariance is restored. As illustrated
in (\ref{eq:8}) the low temperature or low $\tau$ corrections are
exponentially suppressed in $\ha_n$.

The thermal coefficients have been collected according to the mass
dimension of the local operators. At zero temperature this is
equivalent to an expansion in powers of $\tau$, but at finite
temperature it corresponds to expand in powers of $\tau$ keeping
$\tau/\beta^2$ and $\Omega$ constant. Due to the exponential
suppression, a strict expansion in $\tau$ for fixed $\beta$ would
yield just the zero temperature result. (This accounts for the well
known fact that ultraviolet divergences and anomalies are temperature
independent \cite{Dolan:1974qd}.)

Also useful is the expansion of the trace of the heat kernel operator
$\Tr(e^{-\tau K})=\sum_n\tau^n\int d^dx\tr( \hb_n)$. To mass dimension
six we find
\begin{eqnarray}
\hb_0 &=& \varphi_0 \nonumber \\
\hb_1 &=& -\varphi_0 M
\nonumber \\ 
\hb_2 &=& 
\frac{1}{2}\varphi_0 M^2
-\frac{1}{3}\varphi_2 E_i^2 
+\frac{1}{12} \varphi_0 F_{ij}^2
\nonumber \\
\hb_{5/2} &=& 
-\frac{1}{6}\varphi_1 \{ M_i,E_i \} 
 \\
\hb_3 &=& 
-\frac{1}{6}\varphi_0 M^3
+\frac{1}{6}\varphi_2 M_0^2
-\frac{1}{12}\varphi_0 M_i^2
\nonumber \\ &&
+\frac{1}{3}\varphi_2 E_i M E_i
-\left(\frac{1}{12}\varphi_0
+\frac{1}{3}\varphi_2
+\frac{2}{15}\varphi_4\right) E_{0i}^2
\nonumber \\ &&
+\frac{1}{30}\varphi_2 E_{ii}^2
+\frac{1}{30}\varphi_2 F_{0ij}^2
-\frac{1}{15}\varphi_2 E_i F_{ij} E_j
\nonumber \\ &&
-\frac{1}{12}\varphi_0 F_{ij} M F_{ij}
-\frac{1}{60}\varphi_0 F_{iij}^2
+\frac{1}{90}\varphi_0 F_{ij}F_{jk}F_{ki} \,.
\nonumber 
\end{eqnarray}
As is well known, the expression for $\hb_n$ is intrinsically not
unique due to integration by parts and the trace cyclic property. At
zero temperature these operations do not mix different orders, but at
finite temperature this is no longer true since the commutators (in
particular covariant derivatives) of $\Omega(x)$ generate dimensional
terms \cite{Megias:2003prep}. This explains why, e.g. $\hb_{3/2}$
vanishes although $\Tr(\ha_{3/2})$ is not identically zero. Of course,
the trace of the heat kernel itself is unique and the ambiguity
amounts to a reorganization of the asymptotic expansion (no such
ambiguity exists for the $\ha_n$). The heat kernel is symmetric under
transposition $ABC\cdots \to \cdots CBA$ and this symmetry is manifest
in the $\hb_n$ (using the cyclic property). It can be noted that,
exactly as in the zero temperature case \cite{Ball:1989xg}, the
coefficients $\ha_n$ can be obtained from the $\hb_n$ through
functional derivation with respect to $M(x)$.

In addition to the explicit formulas, one of the interesting results
of this calculation is of qualitative type, namely, the operators
appearing in the coefficients $\ha_n$ or $\hb_n$ can be grouped
according to the weights $\varphi_k$. This indicates a definite
pattern of Lorentz symmetry breaking at finite temperature (even in
the absence of scalar potential). This pattern will manifest itself in
applications such as those to renormalization group flow at finite
temperature \cite{Schaefer:1999em}, the general form of the chiral
perturbation theory effective Lagrangian at finite temperature, etc.

The fundamental role played by the Polyakov loop is further
illustrated by considering the trivially solvable case of a
homogeneous relativistic quantum gas. In the absence of external
fields (other than a mass $m$), the lowest order term
$\hb_0=\varphi_0$ is sufficient to obtain the effective action and so
the grand canonical potential ($e^{-\tau m^2}$ exactly factorizes out
of the heat kernel)\footnote{The effects of external fields can be
included through higher orders in the heat kernel expansion.}. A
simple integral transform to pass from the exponential to the
logarithm function reproduces the standard result, with distribution
function
$(e^{\beta(\omega_k-\mu)}-1)^{-1}-(e^{\beta(\omega_k+\mu)}-1)^{-1}$
(for bosons) \cite{Haber:1981fg}. We want to emphasize the relation
between the chemical potential $\mu$ and the Polyakov loop
$\Omega$. The chemical potential couples as a constant additive term
in the scalar potential. Because it is constant, $\mu$ does not
contribute to the local operators, since there $A_0(x)$ only appears
through the covariant derivative $\D_0$. If the Polyakov loop were
absent in the formulas, $\mu$ would not appear at all in the partition
function. An obviously incorrect result. It is also noteworthy that
the periodic dependence of the heat kernel on $\log\Omega$ implies the
well known fact that the partition function is periodic in $\beta\mu$
with period $2\pi i$ (a consistency condition due to its coupling to
the quantized charge operator). The Polyakov loop thus appears as a
generalization of the factor $e^{\beta\mu}$ to non constant and non
Abelian gauge fields. In curved space-times, in addition to the
Polyakov loop of the gauge connection $A_\mu$, there is a Polyakov
loop tied to the parallel transport connection $\Gamma_\mu$, with
consequences in field theory in presence of gravitational fields.

\begin{acknowledgments}
This work is supported in part by funds provided by the Spanish DGI
with grant no. BFM2002-03218, and Junta de
Andaluc\'{\i}a grant no. FQM-225.
\end{acknowledgments}


\end{document}